\documentclass[aps,reprint,prl,superscriptaddress]{revtex4-1}
\usepackage{ulem}
\usepackage{graphicx}
\usepackage{hyperref} 
\pdfoutput=1
\def\ypo4{YPO$_4$}
\def\d#1/d#2{\frac{\partial #1}{\partial #2}}

\def\iu#1{U$^{+#1}$}
\def\ith#1{Th$^{+#1}$}
\def\ix#1{X$^{+#1}$}

\def\rv{\vec{r}}

\usepackage{soul}

\begin{document}
\title{
 Which oxidation state of uranium and thorium as point defects in xenotime is
 favorable?
}
\author{Y.~V. Lomachuk}
\email[]{Lomachuk\_YV@pnpi.nrcki.ru}
\author{D.~A.\ Maltsev}
\author{N.~S.\ Mosyagin}
\author{L.~V.~Skripnikov}
\author{R.~V.~Bogdanov}
\author{A.~V.\ Titov}
\email{Titov\_AV@pnpi.nrcki.ru}
\affiliation{Petersburg Nuclear Physics Institute named by B.P.\ Konstantinov
of National Research Center ``Kurchatov Institute'' (NRC ``Kurchatov
Institute'' - PNPI), 188300, Russian Federation, Leningradskaya oblast,
Gatchina, mkr.\ Orlova roscha, 1.}
\date{\today}
\begin{abstract}
 Relativistic study of xenotime, \ypo4, containing atoms thorium and uranium as point defects is performed in the framework of 
cluster model with using the compound-tunable embedding potential (CTEP) method proposed by us recently \cite{Maltsev:19a}.
The Y--(PO$_4$)$_6$--Y'$_{22}$--O'$_{104}$ cluster for xenotime is considered, in which central part, [Y--(PO$_4$)$_6$]$^{-15}$, is the main cluster, whereas outermost 22 atoms of yttrium and 104 atoms of oxygen are treated as its environment and compose electron-free CTEP with 
the total charge of 
$+15$. 
The P and O atoms of the orthophosphate groups nearest to the central Y atom
are treated at all-electron level. The central Y, its substitutes, Th and U,
together with environmental Y atoms are described within different versions of
the generalized relativistic pseudopotential method \cite{Titov:99}.
Correctness of our cluster and CTEP models, constructed in the paper, is
justified by comparing the Y-O and P-O bond lengths with corresponding periodic
structure values of the \ypo4 crystal, both experimental and theoretical.

Using this cluster model, chemical properties of
  solitary
point defects, X = U, Th, in xenotime are analyzed.
 It has been shown that the oxidation state 
${+3}$ is energetically more profitable
 than 
${+4}$ not only for thorium but for uranium as well ($\Delta E \approx 5$\,eV) despite the notably higher ionic radius of U$^{+3}$ compared to Y$^{+3}$, whereas ionic radii of U$^{+4}$ and Y$^{+3}$ are close. This leads to notable local deformation of crystal 
   geometry 
around the U$^{+3}$ impurity in xenotime and contradicts to widespread
    opinion
about favorite oxidation state of uranium in such kind of minerals \cite{Goldschmidt:54}.
\end{abstract}
\maketitle
\section*{Introduction}
Natural orthophosphates of yttrium and rare earth elements (minerals like xenotime \ypo4 and monazite CePO$_4$) are characterized by high chemical and radiation resistance \cite{Nazdala:18} and are considered as natural analogues of matrices for immobilization of actinides \cite{Dacheux:04, Ji:17, Popa:16}.

Methods for the high-temperature synthesis of ceramics based on these orthophosphates have been developed in detail, conditions for the stabilization of actinides in the tri- and/or tetravalent states have been found \cite{Vance:11, Arinicheva:17, Zhang:08}. (Higher degrees of oxidation of actinides are not formed in this case.) It is assumed that the resulting composites will be buried in deep geological formations for a period of at least 10,000 years \cite{Lumpkin:12}.

An understanding of the immobilization properties of such matrices at the atomic level can be achieved only on the basis of quantum-chemical modeling the electronic structure of the considered actinide-containing materials.

Calculations concerning the electronic structure 
  of solids
are usually carried out taking
into account their periodic structure, 
however, it is reasonable to utilize cluster
models for these minerals due to relatively low
concentration of the impurity actinides in solids.
 
In this case, the calculation is carried out only
for 
the
small region that includes an impurity atom and its environment. The remaining part of the crystal is modeled by an embedding potential, to account for influence of environment on the selected fragment of the crystal. Such a modeling scheme is known as
 the embedded cluster method \cite{Abarenkov:16}.

While good incorporation of Th into monazite is naturally justified 
by approximately the same ionic radii of Ce and Th atoms \cite{Bugaenko:08} 
and similarity of their electronic structure, 
there is no such analogy between Y and U (Th) atoms for the case of xenotime.
Therefore, one needs more detailed consideration based on 
theoretical electronic structure modeling of xenotime with 
the impurity actinide atoms.

The grounds of new, combined approach based on the relativistic study of materials and their fragments with inclusion of impurity atoms (which can be $f$ and heavy $d$ elements)
were recently developed by us based on the compound-tunable embedding potential 
(CTEP) 
method \cite{Maltsev:19a}.
Electronic structure calculations with CTEP of the point defects containing uranium and thorium made it possible to determine
a number of their characteristics.
One of the most interesting questions is that about the energetically preferred 
oxidation state 
of thorium and uranium in xenotime.
    According to
Goldschmidt's long-standing paper 
\cite{Goldschmidt:54} 
it should be +4, not +3.
More recently, Vance {\it et al.} wrote in \cite{Vance:11}
   ``Thus overall the results for U were broadly similar to those for Np and Pu, except that only tetravalent U was observed'' and
 ``U$^{+3}$ should also be able to be incorporated in principle but the necessary conditions would likely be so reducing that the xenotime and monazite structures would be destabilized by the reduction of phosphate to elemental P.''
Thus, the question about 
the +3 oxidation state 
of U in xenotime is yet open and
   one of goals of this research is to
  discuss this problem from theoretical point of view.

  Note, that we consider here only the single-atom point defects, though more complicated substitutes like \hbox{4Y$ \to$ 3U + $\bigcirc_Y$}
(where $\bigcirc_Y$ is a Y--site vacancy, see  \cite{Vance:11} and refs.)
are also possible and will be considered in the other study.

Our cluster model is given in section ``Computational details'', results of calculation are considered in ``Results'', and discussion about absence of experimental data with trivalent U is given in ``Conclusions''.

To summarize, 
the CTEP-based combined approach
developed here to actinide-containing impurities in xenotime
is quite versatile and can be used to study
very different electronic properties of materials with point defects \cite{Maltsev:19a} and various processes, in particular, such as localized vibrations, rotations and electronic excitations in crystals, as well as to study sorption processes of heavy atoms. In paper \cite{Shakhova:19a} it is applied to 
YbF$_2$ and YbF$_3$ 
crystals containing $f$-element, Yb, as regular atom of periodic structure.
\section*{Computational details}
To carry out calculations of the electronic structure of the xenotime 
crystal 
the DFT method with hybrid PBE0 \cite{Adamo:99} functional 
implemented in the {\sc crystal} code \cite{Dovesi:18} was used.
This code allows us to use the same DFT PBE0 functional and basis 
sets as that in 
the cluster model calculations,
    both one- and two-component,
which are discussed 
below. Thus, 
one can directly juxtapose results of these calculations. 

The cluster model calculations were carried out with using the two-component DFT code \cite{Wullen:10}.
For simulating the crystal structure, basis set superposition error (BSSE) arising from presence of diffuse type orbitals is  significant, thus we use the relatively small basis sets for such calculations.
It is also important to note, that using the same basis sets, atomic pseudopotentials (PPs) and DFT-functional for the periodic crystal and cluster model calculations allows one 
    to estimate reliably
the errors arising from using the embedded cluster model simulating
   the crystal fragment.

The PPs for yttrium developed by 
our group \cite{Titov:99, QCPNPI:Basis} was applied 
to exclude core shells from calculation
such that only 11 outermost yttrium electrons are treated explicitly. 
The uncontracted basis set (5s4p3d) \cite{QCPNPI:Basis} was used 
for these electrons.

The oxygen and phosphorus atoms are treated at all-electron level, basis sets for them are taken from \cite{Peintinger:13}.

\begin{figure}
\centering  
\includegraphics[width=0.5\textwidth]{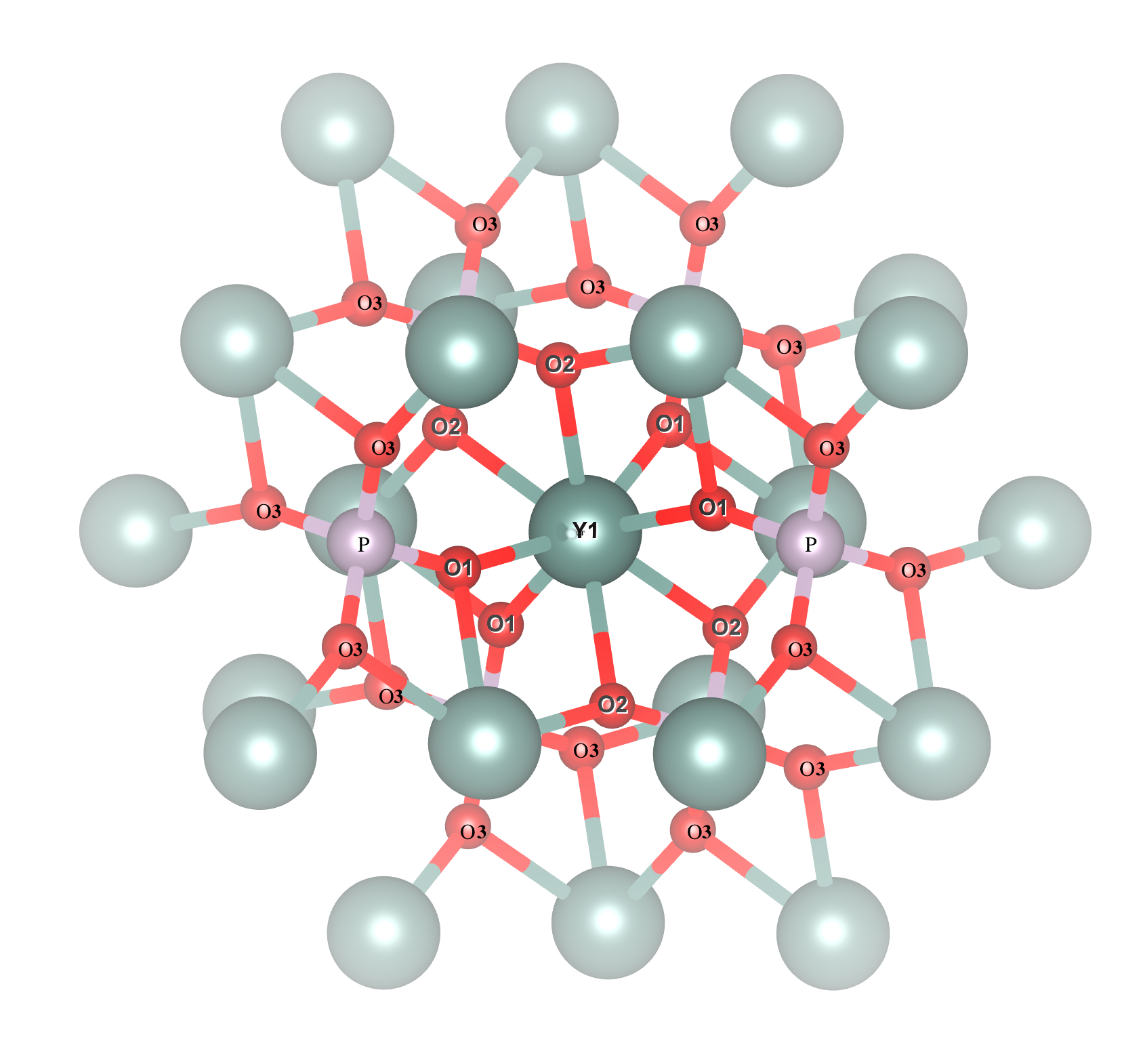}
\caption{Xenotime YPO$_4$ cluster model.
   The cluster model structural formula is \hbox{Y--(PO$_4$)$_6$--Y'$_{22}$--O'$_{104}$}. The anionic layer, O'$_{104}$, is not presented on this figure due to the limited space.
  The cationic layer pseudoatoms, Y'$_{22}$, are displayed as shaded spheres. The central yttrium atom Y1 and P
atoms of main cluster are denoted by corresponding labels. There are three types of O atoms in the main cluster area -- oxygen atoms of these types labeled as O1, O2 and O3.  The Y--O bond lengths between central Y1 atom and oxygen atoms of types O1 and O2 differ from each other. The atoms labeled as O1 belong to the one orthophosphate group in the main cluster area, while atoms of the second type belong to two adjacent orthophosphate groups. The O3 atoms of the main cluster are not chemically bonded to the central yttrium atom.}
\label{fig:xenotime_clust}
\end{figure}

We consider the following cluster model of xenotime crystal to perform 
calculations of properties of Y, Th, U atoms in xenotime: the structural
formula of our cluster is Y-(PO$_4$)$_6$-Y'$_{22}$-O'$_{104}$.  The main cluster, Y-(PO$_4$)$_6$, consists of the central yttrium (substituted later by uranium or thorium) atom and surrounding six orthophosphate groups PO$_4$ (see Figure~\ref{fig:xenotime_clust}).  For the main-cluster atoms, the same pseudopotential and basis sets for Y, P, O atoms were used as for the periodic structure calculations.
 To reproduce the electronic structure of a fragment in a crystal (in particular, saturate  chemical bonds properly) we need first reproduce the oxidation states of the atoms inside the fragment. For the main cluster we have +3 for oxidation state of Y and -3 for each PO$_4$ group.
 Thus, 15 additional electrons should added to the main cluster, which, in turn, should be compensated by the charge of environment (see below).

The cationic layer of the cluster model, Y'$_{22}$, consists of 22 yttrium pseudoatoms, which are modeled with using particular kind of the ``electron free'' pseudopotential \cite{Maltsev:19a}, with respect to the Y$^{+3}$ oxidation state (``electron free'' PP means here that we do not introduce additional electrons to the extended cluster under consideration compared to the main one). The basis sets used for these atoms were taken here the same as for 
the
central yttrium atom. 

The anionic layer, O'$_{104}$, consists of the 104 
   oxygen-site
negative  
point charges without addition of electrons to the cluster as well. The net charge of 15 additional
electrons to the main cluster is completely compensated by the corresponding fractional charges on yttrium (cationic) and oxygen (anionic) sites of the environmental layers. 
In our approach, the additional charges on these 
yttrium 
pseudoatoms as well
as on 
    oxygen-site 
negative point charges of the anion layer are considered as adjustable parameters of the CTEP for xenotime.

In the present work, values of these charges were obtained by minimizing 
root mean square (RMS)
force $|f|$ acting on the atoms of the main cluster. 
This value is calculated as 
\begin{equation} 
\label{eq:av_force} 
|f| = \sqrt{\sum\limits_{i=1}^{N_{at}}\left(\nabla_i E)^2\right/N_{at}},
\end{equation} 
where
$E$ is the evaluated total energy of the cluster, $N_{at}$ -- number of atoms in
the main cluster ($N_{at}=31$ for the xenotime cluster model), and $\nabla_i$ is the gradient 
operator
with respect to coordinates of $i$-th atom.

The basis sets and pseudopotentials developed by our group
\cite{QCPNPI:Basis} were used 
    for calculations of xenotime with the U and Th point defect substitutes of Y.
\section*{results}
\subsection{Periodic structure calculation results}
\def\err#1e-#2{\pm #1\times 10^{-#2}}
\begin{table}[h!]
\caption{Results of the calculations of 
Y-(PO$_4$)$_6$-Y'$_{22}$-O'$_{104}$ 
cluster model and xenotime crystal.}
\label{table:xenotime_clust} 
	\begin{ruledtabular}
\begin{tabular}{lccc}
 &expt. data & crystal    &  cluster  \\
    & & calculations & model calculations\\
\hline\\
$a=b$, \AA\footnotemark[1]  &	6.89  &6.93      &--  \\
$c$, \AA\footnotemark[1]    &	6.03  &6.06  &-- \\
$Q_{Y}$, a.~u.\footnotemark[2]  & &   & $ 2.6 \pm 0.1$\footnotemark[3]  \\
$Q_{O}$, a.~u.\footnotemark[2]  & &   & $-0.4 \pm 0.3$\footnotemark[3]  \\
$|f|$, a.u.\footnotemark[4]     & &   & $3.6\times 10^{-4}$  \\

Y--O1, \AA\footnotemark[5]  &$2.32$    &$2.32$   & $2.317 \pm 0.002$   \\
Y--O2, \AA\footnotemark[5]  &$2.38$    & $2.37$  & $2.372 \pm 0.004$    \\

P--O, \AA \footnotemark[5] & $1.54$   &$1.57$    & $1.566 \pm 0.001$   \\
\end{tabular}
	\end{ruledtabular}
\footnotetext[1]{The lattice parameters obtained from the DFT PBE0 \cite{Adamo:99} periodic structure calculations using the {\sc crystal} \cite{Dovesi:18} code;
the experimental data are taken from \cite{Ni:95}.}

\footnotetext[2]{The $Q_Y$ and $Q_O$ are average additional charge values on
        the Y' pseudoatoms of cluster cation layer Y'$_{22}$ and on the O' pseudoatoms of
        cluster anion layer, correspondingly. The whole sets of these values
        were optimized to minimize 
  RMS
force acting on the main-cluster
atoms, when the atom nuclei positions obtained from the crystal calculations are
used.}
\footnotetext[3]{Standard deviation of the value $x$ calculated on the sets of
        additional charges of cation layer atoms  for $x = Q_Y$ and of anion layer atoms for
        $x = Q_O$. These values are not zero, due to the fact that symmetry of the 
cluster model is lower than symmetry of the xenotime crystal.  
There is only the one atom 
of the each kind (Y, O, or P) in the unit cell of
xenotime, but there are several non equivalent Y and O  atoms in the cluster
model.
}
\footnotetext[4]{%
 RMS
force acting on the atoms of the main cluster 
        Y-(PO$_4$)$_6$.
This value is calculated from equation~(\ref{eq:av_force}).}
\footnotetext[5]{The P--O, Y--O1 and Y--O2 bond lengths (see Figure
        \ref{fig:xenotime_clust}). In the cluster model calculations column,
        average values for the atoms of the main cluster are represented.
        Errors for them are estimated as corresponding standard deviations. It
follows from provided data that errors associated with the embedded cluster model are much
less then the errors arising from using DFT approximation.}

\end{table}
According to the  experimental data \cite{Ni:95} the crystal
system 
of the xenotime YPO$_4$ is tetragonal (lattice parameters $a=b\ne c$,
$\alpha=\beta=\gamma=90^{\circ}$). The space group is $I 4_1/amd$ and there are 3
non-equivalent atoms Y, P, O in the unit cell. 
The lattice parameters as well as positions of 
the atomic nuclei 
within the unit cell of the crystal were optimized to achieve minimum of the total energy of the system.

The results of the calculation are given in Table~\ref{table:xenotime_clust}. Differences between the theoretical values of lattice parameters and Y--O and P--O bonds lengths and corresponding experimental data \cite{Ni:95} are 
within 0.04 \AA, 
this is typical for used DFT approach.
\subsection{Calculation of CTEP parameters for cluster studies}

Using the calculated periodic structure data one can consider the cluster model of the xenotime Y-(PO$_4$)$_6$-Y'$_{22}$-O'$_{104}$ (see Figure~\ref{fig:xenotime_clust}) to generate CTEP.
Due to the lower symmetry of the cluster, for which only point group can be taken into account as compared with xenotime crystal, the yttrium pseudoatoms of the cationic layer are not equivalent to each other; the same is true for the oxygen-site negative point charges of the anionic layer. This fact leads to dispersion in values of additional charges $Q_Y$ on
yttrium pseudoatoms and $Q_O$ on oxygen pseudoatoms. The results of  charge optimization procedure are listed in Table~\ref{table:xenotime_clust}.

Optimal values of charges $Q_Y$ are
   spread
 around average value $\langle
Q_Y \rangle \approx 
 2.6 \pm 0.1$, 
this value is in qualitative agreement with the corresponding formal charge +3. 
  RMS
force acting on the main-cluster atoms $|f|$ is 
of the order of $10^{-4}$ a.\,u. 

To obtain more illustrative estimate of quality of the described cluster model,
positions of the main-cluster atoms were optimized to achieve minimum of the
total energy of the system. During this optimization process, positions of
   pseudoatoms
from cationic and anionic environmental layers were considered as fixed together with the values of additional charges $Q_Y$ and $Q_O$. 
Difference between the optimized cluster and calculated crystal values for the Y--O and P--O bond lengths
is 
of the order of $10^{-3}$ \AA, 
this value is much less than the difference between the results of crystal calculation and experimental data.

To estimate the reproducibility of the electronic properties of the original non-substituted cluster, compared to the solid-state calculations, electronic density cube files were obtained for the periodic crystal study and for the cluster with CTEP. The cube grid was chosen to be the same in both cases with the orthogonal unit vectors of about 0.053 a.u.

As a quantitative criterion we provide the angle-averaged 
difference between the cluster and crystal electronic densities, 
calculated by the following formula:
\[ d(r) = \frac{1}{4\pi}\oint d \Omega \left | \rho_{cluster}(\vec{r})-\rho_{crystal}(\vec{r}) \right | \ ,\]
\noindent where the 
 $\vec{r}$ is the
radius vector of the point with origin in the central atom nucleus site, the
$d\Omega$ is the differential solid angle, and the $d(r)$ represents the
absolute magnitude of difference.

\begin{figure}
\centering  
\includegraphics[width=0.5\textwidth]{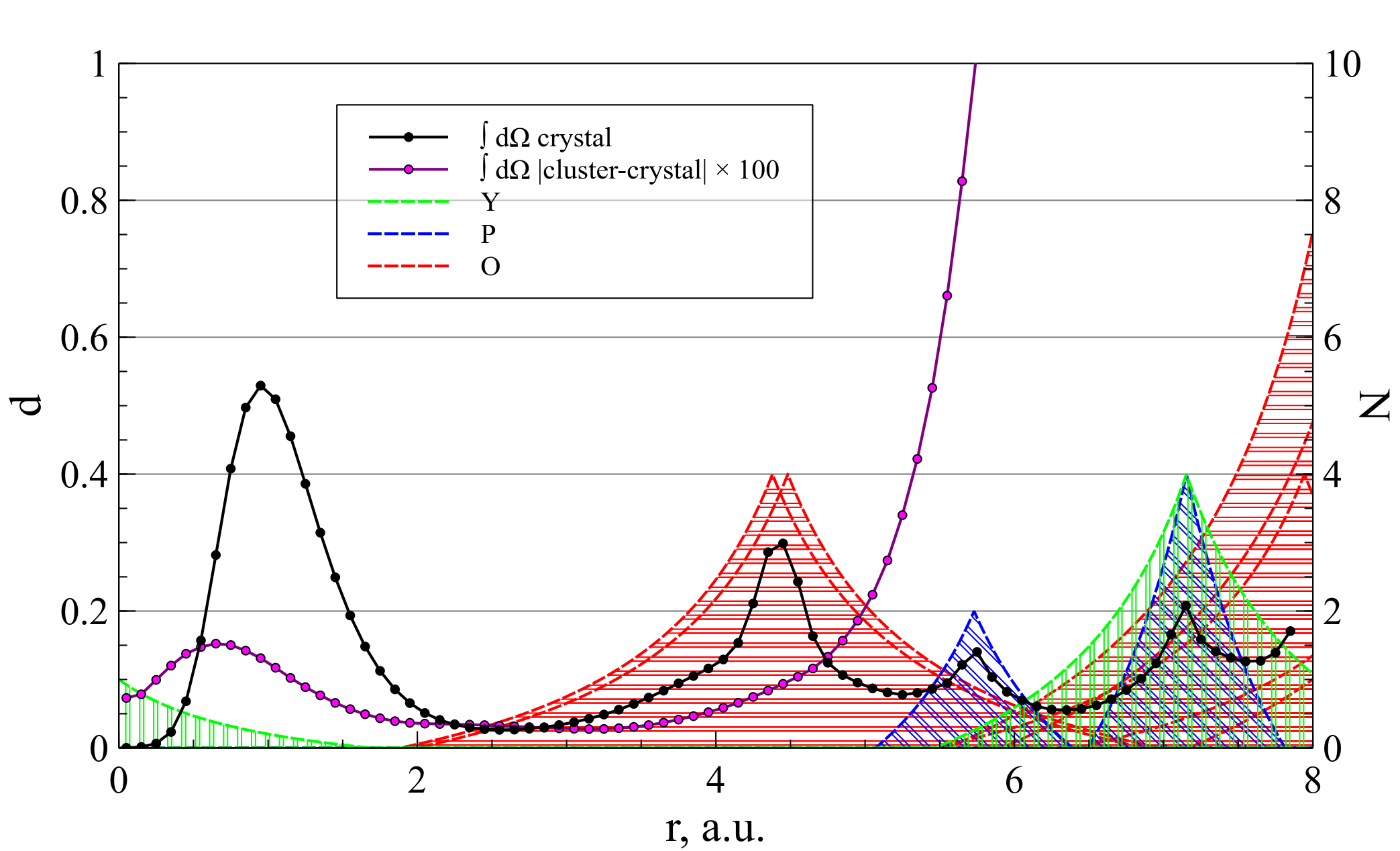}
	\caption{The radial dependence of electronic density differences for the original (non-substituted) cluster. Solid line corresponds to the integral of absolute difference $d(r)$. 
Filled peaks at the bottom qualitatively represent the position of the neighbour atoms (color denotes atom type, width at the bottom is equal to crystal radius, and the peak height is proportional to the number 
(N on the right scale)
of atoms at the same distance to the center).}
\label{fig:dens}
\end{figure}

This value is plotted at the Figure \ref{fig:dens} as the solid line. The
bottom curves qualitatively represent the electronic density of atoms from the
main cluster. The black curve is the total density, and the difference curve is
multiplied by factor 
of 
100.

\subsection{Properties of Th and U in xenotime}

With the above constructed xenotime cluster model, effective states of impurity uranium and thorium atoms in xenotime were studied in calculations of the clusters with the above generated CTEP, in which the central yttrium atom is substituted by the U or Th atom. For both cases four different types of calculations were carried out, with 15, 14, 13 and 12 additional electrons in the main cluster (that is not neutral 
in the last three cases
in contrast to the extended one at the CTEP generation step). 
The first 
two calculations are corresponding to the cases of X$^{+3}$ and X$^{+4}$ (X= Th, U) oxidation states of the appropriate defects in xenotime.
While the clusters with 13 and 12 additional electrons for central uranium atom case correspond to the cases of the \iu5 and \iu6 oxidation states, analogous clusters with central thorium atom correspond to the \ith4 oxidation state and ionized neighboring PO$_4$ groups (see Table~\ref{table:th_u_in_xenotime}).
Positions of atoms in the main cluster with actinides were optimized to minimize its total energy. Results of these calculations are listed in Table~\ref{table:th_u_in_xenotime}.
\def\acs#1p#2d#3f#4e{7s^{#1} 7p^{#2} 6d^{#3} 5f^{#4}}
\def\las#1p#2d#3f#4e{5s^{#1} 4p^{#2} 4d^{#3}}
\def\clust{Y -- (PO$_4$)$_6$-Y'$_{22}$-O'$_{104}$}
\def\nce{ -- (PO$_4$)$_6$-Y'$_{22}$-O'$_{104}$}
\def\ypo4{YPO$_4$\ }
\def\iathn{TPO\ }
\def\iath#1{(TPO)$^{+#1}$\ }
\def\iaun{UPO\ }
\def\iau#1{(UPO)$^{+#1}$\ }
\def\iax#1{(XPO)$^{+#1}$\ }
\begin{table*}[h!]
        \caption{Thorium and uranium atoms in xenotime properties$^a$}
        \label{table:th_u_in_xenotime}
\begin{ruledtabular}
\begin{tabular}{lccccccccc}
        & \ypo4 \footnotemark[1]                                  &  \iathn     & \iath1   &\iath2 & \iath3 & \iaun & \iau1 & \iau2 & \iau3 \\
        $Q_X$, bader charge, a.~u.\footnotemark[2]         &      &	   2.4      &	3.0    &   3.0 & 3.0    &  2.3  & 2.8   & 3.0   & 3.2 \\
        $\langle |\vec{S}| \rangle$, a.~u.\footnotemark[3] &      &   1.1       &   0.0    &   1.2 & 2.4    &  2.8  & 1.9   & 1.1   & 0.0 \\

        P - O, \AA\footnotemark[4]      &   1.56   &    1.56     &	 1.54  --  1.58 & 1.54  --  1.58 & 1.54  --  1.56 & 1.56 & 1.55  --  1.59 & 1.54 -- 1.62 & 1.51 -- 1.67 \\	
        X - O1, \AA\footnotemark[4]     &   2.32   &    2.42	 &   2.32	      & 2.32         & 2.32 & 2.40 & 2.29    & 2.15  -- 2.34 & 2.06  --  2.10 \\ 
        X - O2, \AA                     &   2.37   &    2.47     &	 2.40	      & 2.37         & 2.4  & 2.45 & 2.37    & 2.15  -- 2.34 & 2.25  --  2.29\\ 
        R$_X$, \AA\footnotemark[4]	    &   0.9    &    1.04	 &   0.94         &  &  & 1.00 & 0.89    & & \\

$\Delta E_{ion}$ , eV \footnotemark[5]   &  --  &   -- & 3.5 & 18.0 & 35.0 &  --  &6.2 & 11.2 & 25.0\\
\end{tabular}
\end{ruledtabular}
\footnotetext[1]{Results of the two-component DFT PBE0 calculations of the cluster models.
        The \ypo4 column coressponds to the xenotime cluster model \clust,
        \iathn and \iaun correspond to the X\nce, X = Th, U clusters, columns
        \iax{n}, $n=1,2,3$, X=Th, U correspond to the ionized clusters. 
}
\footnotetext[2]{The Bader net charge values of the central atom $Q_X$ (X=Y,
        Th, U) are obtained with using the computer code \cite{Tang:09}. It
        worth mentioning that the  bader charges of Th and U ions with equal
        formal charges also approximately equal.  The equal Thorium ion charge values for
        the cases of \iath1, \iath2 and \iath3
        clusters prove that 
ionization of the PO$_4$ group
        occurs instead of further ionization of the central atom 
in the former two cases.
}
\footnotetext[3]{
The value of total spin for the system, 
calculated as $\langle |\vec{S}| \rangle =
\int \sqrt{S_x(\rv)^2 +S_y(\rv)^2 + S_z(\rv)^2}dV$.  For the spin density
distribution in the considered clusters, see
Fig~\ref{fig:sd_u} and  Fig~\ref{fig:sd_th}. The non-zero values of spin
of the clusters \iath2 and \iath3 correspond to the unpaired electrons on
the PO$_4$ groups in these cases.}

\footnotetext[4]{Lengths of the P -- O and X -- O bonds. 
        Ionic radii
        $\mathrm{R}_X$ of the \ith4 and \iu4 (see 
paper \cite{Bugaenko:08} 
and row $R_X$ of
        this table)  are approximately equal to that of the Y$^{+3}$ cation.
        This statement 
agrees 
with that the \iu4, \ith4 substitutes incapsulated  in
        the crystal deform their nearest environment in much less extent than
        \iu3 and \ith3 substitutes.
	}
\footnotetext[5]{Energy of the cluster ionization $\mathrm{XPO}\to
        (\mathrm{XPO})^{+n}$, $n=1, 2, 3$, X = U, Th  calculated as difference between
total energies of the corresponding systems.}
\end{table*}

\begin{figure}
\centering  
\includegraphics[width=0.5\textwidth]{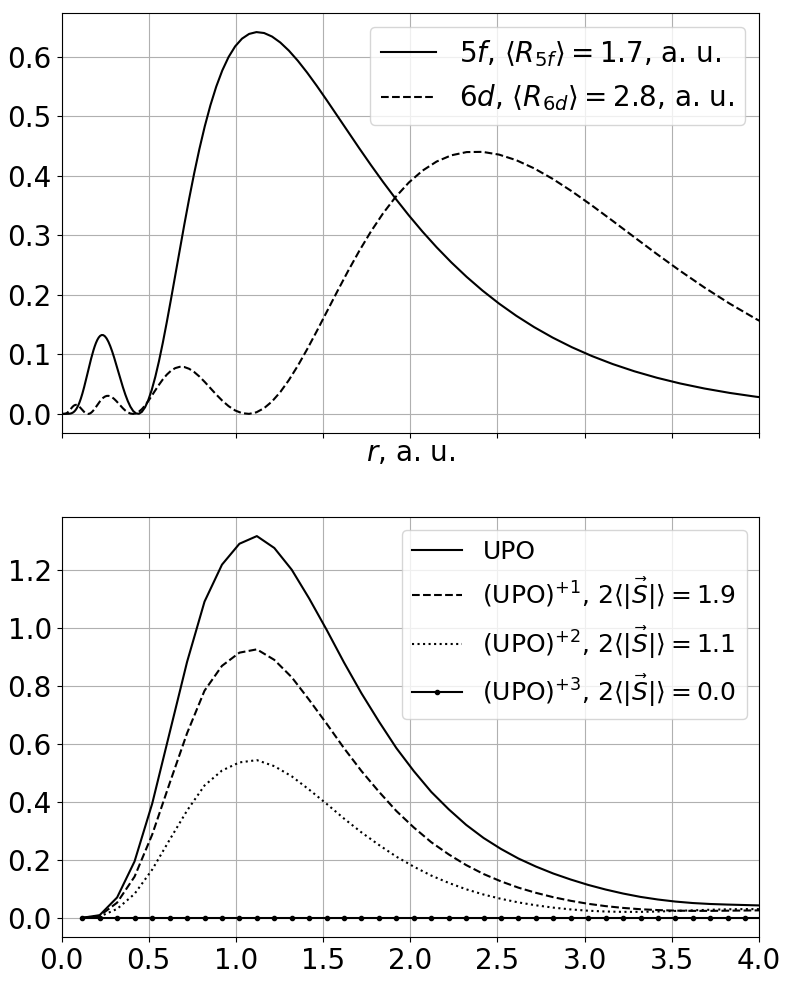}
	\caption{Radial electron densities corresponding to the $5f$ and $6d$
	one-electron states of the uranium 
neutral 
free atom calculated with {\sc hfd}
	code \cite{hfd} (at top) and  spin density \hbox{$|\vec{S}(r)| = \sqrt{S_x^2(r) + S_y^2(r) + S_z^2(r)}$} distribution as function of the
	distance from central uranium atom in the
	\hbox{U$-(\mathrm{PO}_4)_6
-
\mathrm{Y'}_{22}-\mathrm{O'}_{104}$}
	clusters (at bottom). The spin density distributions of the clusters
	with 15, 14, 13 and 12 additional electrons (denoted as \iaun, \iau1, \iau2 and \iau3,
	correspondingly) is almost proportional to each other. It follows
	from the presented data that total spin values of the clusters arise
	from different numbers of electrons on the open $5f$ shell of U.
	 These clusters model different oxidation  states of uranium in xenotime.
}
\label{fig:sd_u}
\end{figure}

\begin{figure}
\centering  
\includegraphics[width=0.5\textwidth]{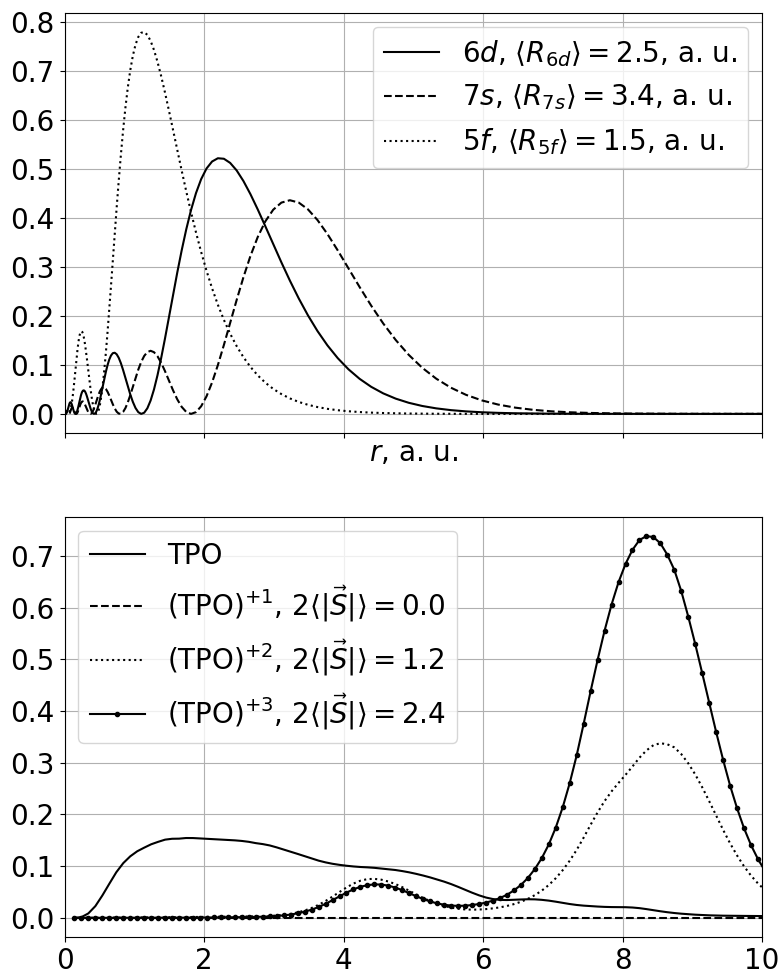}
	\caption{Radial electron densities corresponding to the $6d$ and $7s$
    one-electron states of the free thorium ion with electronic configuration
    $[Rn]7s^{0.3} 7p^{0.3} 6d^{0.3} 5f^{0.1}$ calculated with {\sc hfd}
	code \cite{hfd} (at top) and  spin density \hbox{$|\vec{S}(r)| = \sqrt{S_x^2(r) + S_y^2(r) + S_z^2(r)}$}
	distribution as function of  the
	distance from central thorium atom in the
	\hbox{Th$-(\mathrm{PO}_4)_6-\mathrm{Y'}_{22}-\mathrm{O'}_{104}$}
	clusters (at bottom). 
The spin density distributions of the clusters
	with 15, 14, 13 and 12 additional electrons (denoted as \iathn, \iath1,
	\iath2 and \iath3 correspondingly) are presented. From these data one can
	conclude that there is one electron on the thorium valence shells in
	the cluster with 15 additional electrons (this case corresponds to the
    $6d^x7s^{1-x}$ configuration of the free Th$^{3+}$ ion) and no electrons on
	these shells in the clusters with 14 additional electrons (this case
	corresponds to the Th$+4$ oxidation state). The presence of the spin density peak at
	8 a.~u. shows that the non-zero total spin value of the clusters with
	13 and 12 additional electrons arise from unnatural ionization of the
	PO$_4$ groups.} 
\label{fig:sd_th}
\end{figure}

To determine if the cases of 14, 13 and 12 additional electrons in the main cluster correspond to the  X$^{+4}$, \ix5, \ix6 oxidation states of the central atom X (Th or U), or not, Bader's charge analysis \cite{Bader:98} were performed for all 
the studied clusters with using {\sc bader} code \cite{Tang:09}. The evaluated Bader net charges show that the case of 14 additional electrons corresponds to the X$^{+4}$ substitute in the main cluster (X=Th, U), whereas 13 and 12 additional electrons in the main cluster correspond to 
X$^{+5}$ and X$^{+6}$ substitute only for case of X=U. 
Additionally, we present the spin density distribution in the clusters on Figures \ref{fig:sd_u} and \ref{fig:sd_th}. It follows from the data that numbers of electrons on the open $5f$-shell of U
correlate with the numbers of additional electrons in the cluster. Note that one cannot extract the corresponding information from the conventional population or Bader analyses since they 
take into account 
contribution of the $f$-orbitals to U-O bonding states that largely compensate (more than on 50\%) the change in the number of open $f$-shell electrons in   the clusters with different number of additional electrons.

There are experimentally measurable properties of atom in compound \cite{Titov:14a} that correlate with its oxidation state.
 First of all they are the x-ray emission spectra chemical shifts of $K\alpha_{1,2}$ lines (transitions $2p_{3/2} \to 1s$, $2p_{1/2} \to 1s$, correspondingly, see \cite{Lomachuk:13, Lomachuk:18en}).
We estimated values of the chemical shifts of the $K\alpha_{1,2}$ lines
 of thorium and uranium substitutes in xenotime with respect to the corresponding free ion by the method described in papers \cite{Titov:14a, Lomachuk:13}. These data are presented in Table~\ref{table:th_u_xes_xenotime}.

The evaluated Y--O and P--O bond lengths shows that the defects 
Th and U 
 being in 
the +3 oxidation state
 deform crystal cell much more than
those in 
the +4 oxidation state.
 This can be explained by the fact, that despite the trivalent Th and U have the same formal
charge as Y$^{+3}$ in YPO$_4$, the tetravalent ones have significantly smaller ionic radii \cite{Bugaenko:08}, which are comparable with that of Y$^{+3}$.

The $K\alpha_{1,2}$ chemical shifts are evaluated for the cluster models with 14, 13 and 12 additional electrons in the considered clusters. For the cases corresponding to different oxidation states of U and Th cations in xenotime, they differ from each other by 
$100 \div 200$ 
meV, while the difference between Th-centered clusters with the same oxidation state of thorium is almost zero. 


\def\acs#1p#2d#3f#4e{7s^{#1} 7p^{#2} 6d^{#3} 5f^{#4}}
\def\las#1p#2d#3f#4e{5s^{#1} 4p^{#2} 4d^{#3}}
\def\clust{Y -- (PO$_4$)$_6$-Y'$_{22}$-O'$_{104}$}
\def\nce{ -- (PO$_4$)$_6$-Y'$_{22}$-O'$_{104}$}
\def\ypo4{YPO$_4$\ }
\def\iathn{TPO\ }
\def\iaun{UPO\ }
\begin{table*}[h!]
	\caption{ Uranium and Thorium x-ray emission spectra $K\alpha_{1,2}$ 
lines
chemical shifts in xenotime cluster calculations.}
\label{table:th_u_xes_xenotime}
\begin{ruledtabular}
\begin{tabular}{lccccccc}
        &U$^{+4}$\footnotemark[1]       & U$^{3+}$ & \iaun\footnotemark[2]& \iau1   & \iau2 & \iau3 \\
	$\chi_{2p1/2\to 1s1/2}$, meV \footnotemark[3] & 0             &  259     &  189            & -10                       & -174              & -285             \\
	$\chi_{2p3/2\to 1s1/2}$,  meV \footnotemark[3] & 0             &  332    &  240             & -15                      & -227              & -372             \\
			     &Th$^{+4}$ \footnotemark[1]      & Th$^{3+}$  & \iathn \footnotemark[2]  & \iath1   & \iath2  & \iath3 \\
	$\chi_{2p1/2\to 1s1/2}$, meV \footnotemark[3]& 0             &  265        &  869                & 798                                   & 796              & 795             \\
	$\chi_{2p3/2\to 1s1/2}$, meV \footnotemark[3]& 0             &  338        &  461                & 372                                   & 370              & 369             \\
\end{tabular}
\end{ruledtabular}
\footnotetext[1]{Free $U^{3+}$, U$^{4+}$, Th$^{3+}$, Th$^{4+}$ ions 
with the
electronic configurations calculated using the two-component DFT PBE0 framework.}
\footnotetext[2]{Cluster model calculations of Uranium and Thorium atoms in xenotime (see the Table \ref{table:th_u_in_xenotime} and the text of the paper for details.)}
\footnotetext[3]{The chemical shifts values of the energies of Uranium and
        Thorium $K\alpha_{1,2}$ transitions in xenotime with respect to the corresponding X$^{+4}$ free ion are obtained from the results of electronic
structure calculations by the method described in works \cite{Titov:14a,
Lomachuk:13}. It is follows from the presented data, that chemical shifts
of 
values of energies of these transitions correlate with the formal charges of
corresponding Uranium and Thorium admixture ions in the xenotime, spin density
distributions presented on the Fig.~\ref{fig:sd_u} and Fig.~\ref{fig:sd_th}, and Bader net charge values listed in the Tabl.~\ref{table:th_u_in_xenotime}.}
\end{table*}
\section*{conclusions}
 Results of the embedded cluster calculations of properties of point
defects in xenotime containing Th and U atoms are presented and discussed. The electronic structure studies are performed using hybrid DFT functional, PBE0 \cite{Adamo:99}, and different versions of the generalized relativistic pseudopotential theory \cite{Titov:99}. The cluster model Y--(PO$_4$)$_6$--Y'$_{22}$--O'$_{104}$ for xenotime YPO$_4$ was used.

The correctness of this model is justified by comparing the optimal Y-O and P-O bond lengths obtained from the cluster calculations and corresponding values from the \ypo4 periodic study. Differences between the bond lengths obtained from the cluster model and periodic crystal calculations are 
of the order of $0.001$ \AA\ 
and are much smaller than differences between results of the crystal structure calculations and experimental data, which are about $0.04$ \AA.  One can conclude that the errors, arising from using the cluster model  are smaller by order of magnitude than those arising from using the DFT PBE0 approximation.

A good agreement of results of cluster modeling and periodic structure
calculations for xenotime shows that 
the
suggested cluster model with CTEP provides reliable data on the total energy as function of coordinates for the main-cluster atoms.
This leads to the possibility of studying embedding of uranium and thorium atoms into the crystal in the framework of 
the
suggested cluster model.

Using this model, the properties of X = U, Th in xenotime were calculated. It
has been shown that the oxidation states X(III) are energetically more
favorable than X(IV)( $\Delta E \approx 5$ eV; to avoid
misleadings arabian designation of both ionic and oxidation states for U and
Th, we use below only roman designations for their oxidation states). 

The x-ray emission spectra chemical shifts of $K\alpha_{1,2}$ lines of Th and U in xenotime compared to the free Th$^{4+}$, U$^{4+}$ ions were calculated within the cluster models corresponding to the U(III), U(IV), U(V), U(VI)
as well as Th(III) and Th(IV) oxidation states of the actinide cations in xenotime. 
The chemical shifts values correlate with the formal charge of the cations, Bader net charges and spin density cluster distributions. 
Both \hbox{Th--(PO$_4$)$_6$--Y'$_{22}$--O'$_{104}$} cluster models with
13 and 12 additional electrons describe Th(IV) oxidation state
in different environments, accordingly; 
the thorium $K\alpha_{1,2}$ 
lines
chemical shifts in these cluster models almost equal to each other.

The obtained result about energetically most profitable oxidation state uranium(III) is in somewhat contradiction with only observed oxidation state U(IV) \cite{Vance:11}.
  As is shown here, substitution of
        U(III) instead of Y(III) 
  does not lead to local geometry perturbation of the U-neighboring orthophosphate groups; they are rather shifted and rotated  as a whole (xenotime-to-monazite like structure transformation) compared to the central atom.
    Authors of \cite{Vance:11} explained 
   unobservability of U(III)
by too reducing conditions to form trivalent U in xenotime (see ``Introduction''). However, why the trivalent uranium was also not found in single-atom point defects in  natural xenotime, i.e.\ after geologic-scale storage time?

    Our calculations show (see Table~\ref{table:th_u_in_xenotime}) that ionization of {\it any} electron in the U(III)--(PO$_4$)$_6$ fragment leads to its [U(IV)--(PO$_4$)$_6$]$^{+}$ state (note, that ionizing radiation is inherent to actinide-containing minerals). In turn, electron attachment (EA) to the orthophosphate groups in [U(IV)--(PO$_4$)$_6$]$^{+}$ can hardly be suggested as expected (LUMO energy for orthophosphate group is positive) and only direct EA to vacant $5f$ states of U(IV) is energetically profitable ($5f$-LUMO energy of U(IV) is 
${\approx}-0.2$ a.~u.) 
but its ionic radius (${\sim}5f$-orbital size), that is only ${\sim}0.1$~nm compared to that of whole main cluster, ${\sim}0.6$~nm, and high angular momentum ($l{=}3$) dramatically reduce such a probability.
 So, we expect that ionization
 of U(III)-in-xenotime is much
    more likely
 than
    the
electron affinity in U(IV)-in-xenotime. Add here that ambient-electron
      transfer
 through the orthophosphate barrier from other defects is also unlikely.
        In turn, ionization of U(IV)-in-xenotime is not profitable compared to the electron affinity to U(V)-in-xenotime since  the total Bader charge on neighboring orthophosphate groups, (PO$_4$)$_6$, is about 1.3 a.u.\ smaller for U(V)-in-xenotime than that for the case of lowest-energy U(III)-in-xenotime as one can see from Table~\ref{table:th_u_in_xenotime}. As a consequence, the (PO$_4$)$_6$ group around U(V) has a positive electron affinity in our eatimates.
The detailed radiation analysis of actinide-containing xenotime is not the subject of present quantum-chemical research, it requires particular consideration elsewhere.
     The other reason for unobservability of U(III)-in-xenotime is that substitution of neighboring atoms by those with smaller oxidation state takes place in natural xenotime
      since it may also contain minor Ca(II) on Y site, Si(IV) on P(V) site, F(I) on O site and other elements, which can dramatically change the relative profitability of U(IV) vs.\ U(III) as is estimated here.

\section*{Acknowledgements}
Calculations in the paper were carried out using resources of the collective usage centre ``Modeling and predicting properties of materials'' at NRC ``Kurchatov Institute'' - PNPI.
We are grateful to Prof.\ C.\ van~Wullen for the code of modeling the
electronic structure with the use of two-component DFT version \cite{Wullen:10} and to Demidov Yu.~A. for the provided yttrium basis set.
This study was supported by the Russian Science Foundation (Grant No.~14-31-00022).


\begin{thebibliography}{27}%
\makeatletter
\providecommand \@ifxundefined [1]{%
 \@ifx{#1\undefined}
}%
\providecommand \@ifnum [1]{%
 \ifnum #1\expandafter \@firstoftwo
 \else \expandafter \@secondoftwo
 \fi
}%
\providecommand \@ifx [1]{%
 \ifx #1\expandafter \@firstoftwo
 \else \expandafter \@secondoftwo
 \fi
}%
\providecommand \natexlab [1]{#1}%
\providecommand \enquote  [1]{``#1''}%
\providecommand \bibnamefont  [1]{#1}%
\providecommand \bibfnamefont [1]{#1}%
\providecommand \citenamefont [1]{#1}%
\providecommand \href@noop [0]{\@secondoftwo}%
\providecommand \href [0]{\begingroup \@sanitize@url \@href}%
\providecommand \@href[1]{\@@startlink{#1}\@@href}%
\providecommand \@@href[1]{\endgroup#1\@@endlink}%
\providecommand \@sanitize@url [0]{\catcode `\\12\catcode `\$12\catcode
  `\&12\catcode `\#12\catcode `\^12\catcode `\_12\catcode `\%12\relax}%
\providecommand \@@startlink[1]{}%
\providecommand \@@endlink[0]{}%
\providecommand \url  [0]{\begingroup\@sanitize@url \@url }%
\providecommand \@url [1]{\endgroup\@href {#1}{\urlprefix }}%
\providecommand \urlprefix  [0]{URL }%
\providecommand \Eprint [0]{\href }%
\providecommand \doibase [0]{http://dx.doi.org/}%
\providecommand \selectlanguage [0]{\@gobble}%
\providecommand \bibinfo  [0]{\@secondoftwo}%
\providecommand \bibfield  [0]{\@secondoftwo}%
\providecommand \translation [1]{[#1]}%
\providecommand \BibitemOpen [0]{}%
\providecommand \bibitemStop [0]{}%
\providecommand \bibitemNoStop [0]{.\EOS\space}%
\providecommand \EOS [0]{\spacefactor3000\relax}%
\providecommand \BibitemShut  [1]{\csname bibitem#1\endcsname}%
\let\auto@bib@innerbib\@empty
\bibitem [{\citenamefont {Maltsev}\ \emph {et~al.}(2019)\citenamefont
  {Maltsev}, \citenamefont {Lomachuk}, \citenamefont {Shakhova}, \citenamefont
  {Mosyagin}, \citenamefont {Skripnikov},\ and\ \citenamefont
  {Titov}}]{Maltsev:19a}%
  \BibitemOpen
  \bibfield  {author} {\bibinfo {author} {\bibfnamefont {D.~A.}\ \bibnamefont
  {Maltsev}}, \bibinfo {author} {\bibfnamefont {Y.~V.}\ \bibnamefont
  {Lomachuk}}, \bibinfo {author} {\bibfnamefont {V.~M.}\ \bibnamefont
  {Shakhova}}, \bibinfo {author} {\bibfnamefont {N.~S.}\ \bibnamefont
  {Mosyagin}}, \bibinfo {author} {\bibfnamefont {L.~V.}\ \bibnamefont
  {Skripnikov}}, \ and\ \bibinfo {author} {\bibfnamefont {A.~V.}\ \bibnamefont
  {Titov}},\ }\href@noop {} {\  (\bibinfo {year} {2019})},\ \bibinfo {note} {to
  be published}\BibitemShut {NoStop}%
\bibitem [{\citenamefont {Titov}\ and\ \citenamefont
  {Mosyagin}(1999)}]{Titov:99}%
  \BibitemOpen
  \bibfield  {author} {\bibinfo {author} {\bibfnamefont {A.~V.}\ \bibnamefont
  {Titov}}\ and\ \bibinfo {author} {\bibfnamefont {N.~S.}\ \bibnamefont
  {Mosyagin}},\ }\href@noop {} {\bibfield  {journal} {\bibinfo  {journal}
  {Int.\ J.\ Quantum Chem.}\ }\textbf {\bibinfo {volume} {71}},\ \bibinfo
  {pages} {359} (\bibinfo {year} {1999})}\BibitemShut {NoStop}%
\bibitem [{\citenamefont {M.}(1954)}]{Goldschmidt:54}%
  \BibitemOpen
  \bibfield  {author} {\bibinfo {author} {\bibfnamefont {G.~V.}\ \bibnamefont
  {M.}},\ }\href@noop {} {\emph {\bibinfo {title} {Geochemistry}}}\ (\bibinfo
  {publisher} {Oxford, Clarendon Press},\ \bibinfo {year} {1954})\ p.\ \bibinfo
  {pages} {730}\BibitemShut {NoStop}%
\bibitem [{\citenamefont {Nasdala}\ \emph {et~al.}(2018)\citenamefont
  {Nasdala}, \citenamefont {Akhmadaliev}, \citenamefont {Artac}, \citenamefont
  {Chanmuang~N.}, \citenamefont {Habler},\ and\ \citenamefont
  {Lenz}}]{Nazdala:18}%
  \BibitemOpen
  \bibfield  {author} {\bibinfo {author} {\bibfnamefont {L.}~\bibnamefont
  {Nasdala}}, \bibinfo {author} {\bibfnamefont {S.}~\bibnamefont
  {Akhmadaliev}}, \bibinfo {author} {\bibfnamefont {A.}~\bibnamefont {Artac}},
  \bibinfo {author} {\bibfnamefont {C.}~\bibnamefont {Chanmuang~N.}}, \bibinfo
  {author} {\bibfnamefont {G.}~\bibnamefont {Habler}}, \ and\ \bibinfo {author}
  {\bibfnamefont {C.}~\bibnamefont {Lenz}},\ }\href@noop {} {\bibfield
  {journal} {\bibinfo  {journal} {Physics and Chemistry of Minerals}\ }\textbf
  {\bibinfo {volume} {45}},\ \bibinfo {pages} {855} (\bibinfo {year}
  {2018})}\BibitemShut {NoStop}%
\bibitem [{\citenamefont {Dacheux}\ \emph {et~al.}(2004)\citenamefont
  {Dacheux}, \citenamefont {Clavier}, \citenamefont {Robisson}, \citenamefont
  {Terra}, \citenamefont {Audubert}, \citenamefont {Lartigue},\ and\
  \citenamefont {Guy}}]{Dacheux:04}%
  \BibitemOpen
  \bibfield  {author} {\bibinfo {author} {\bibfnamefont {N.}~\bibnamefont
  {Dacheux}}, \bibinfo {author} {\bibfnamefont {N.}~\bibnamefont {Clavier}},
  \bibinfo {author} {\bibfnamefont {A.-C.}\ \bibnamefont {Robisson}}, \bibinfo
  {author} {\bibfnamefont {O.}~\bibnamefont {Terra}}, \bibinfo {author}
  {\bibfnamefont {F.}~\bibnamefont {Audubert}}, \bibinfo {author}
  {\bibfnamefont {J.-E.}\ \bibnamefont {Lartigue}}, \ and\ \bibinfo {author}
  {\bibfnamefont {C.}~\bibnamefont {Guy}},\ }\href@noop {} {\bibfield
  {journal} {\bibinfo  {journal} {Comptes Rendus Chimie}\ }\textbf {\bibinfo
  {volume} {7}},\ \bibinfo {pages} {1141} (\bibinfo {year} {2004})}\BibitemShut
  {NoStop}%
\bibitem [{\citenamefont {Ji}\ \emph {et~al.}(2017)\citenamefont {Ji},
  \citenamefont {Beridze}, \citenamefont {Bosbach},\ and\ \citenamefont
  {Kowalski}}]{Ji:17}%
  \BibitemOpen
  \bibfield  {author} {\bibinfo {author} {\bibfnamefont {Y.}~\bibnamefont
  {Ji}}, \bibinfo {author} {\bibfnamefont {G.}~\bibnamefont {Beridze}},
  \bibinfo {author} {\bibfnamefont {D.}~\bibnamefont {Bosbach}}, \ and\
  \bibinfo {author} {\bibfnamefont {P.~M.}\ \bibnamefont {Kowalski}},\
  }\href@noop {} {\bibfield  {journal} {\bibinfo  {journal} {Journal of Nuclear
  Materials}\ }\textbf {\bibinfo {volume} {494}},\ \bibinfo {pages} {172 }
  (\bibinfo {year} {2017})}\BibitemShut {NoStop}%
\bibitem [{\citenamefont {Popa}\ \emph {et~al.}(2016)\citenamefont {Popa},
  \citenamefont {Cologna}, \citenamefont {Martel}, \citenamefont {Staicu},
  \citenamefont {Cambriani}, \citenamefont {Ernstberger}, \citenamefont
  {Raison},\ and\ \citenamefont {Somers}}]{Popa:16}%
  \BibitemOpen
  \bibfield  {author} {\bibinfo {author} {\bibfnamefont {K.}~\bibnamefont
  {Popa}}, \bibinfo {author} {\bibfnamefont {M.}~\bibnamefont {Cologna}},
  \bibinfo {author} {\bibfnamefont {L.}~\bibnamefont {Martel}}, \bibinfo
  {author} {\bibfnamefont {D.}~\bibnamefont {Staicu}}, \bibinfo {author}
  {\bibfnamefont {A.}~\bibnamefont {Cambriani}}, \bibinfo {author}
  {\bibfnamefont {M.}~\bibnamefont {Ernstberger}}, \bibinfo {author}
  {\bibfnamefont {P.~E.}\ \bibnamefont {Raison}}, \ and\ \bibinfo {author}
  {\bibfnamefont {J.}~\bibnamefont {Somers}},\ }\href@noop {} {\bibfield
  {journal} {\bibinfo  {journal} {Journal of the European Ceramic Society}\
  }\textbf {\bibinfo {volume} {36}},\ \bibinfo {pages} {4115 } (\bibinfo {year}
  {2016})}\BibitemShut {NoStop}%
\bibitem [{\citenamefont {Vance}\ \emph {et~al.}(2011)\citenamefont {Vance},
  \citenamefont {Zhang}, \citenamefont {McLeod},\ and\ \citenamefont
  {Davis}}]{Vance:11}%
  \BibitemOpen
  \bibfield  {author} {\bibinfo {author} {\bibfnamefont {E.~R.}\ \bibnamefont
  {Vance}}, \bibinfo {author} {\bibfnamefont {Y.}~\bibnamefont {Zhang}},
  \bibinfo {author} {\bibfnamefont {T.}~\bibnamefont {McLeod}}, \ and\ \bibinfo
  {author} {\bibfnamefont {J.}~\bibnamefont {Davis}},\ }\href {\doibase
  DOI:101016/jjnucmat201012241} {\bibfield  {journal} {\bibinfo  {journal}
  {Journal of Nuclear Materials}\ }\textbf {\bibinfo {volume} {409}},\ \bibinfo
  {pages} {221} (\bibinfo {year} {2011})}\BibitemShut {NoStop}%
\bibitem [{\citenamefont {Arinicheva}\ \emph {et~al.}(2017)\citenamefont
  {Arinicheva}, \citenamefont {Popa}, \citenamefont {Scheinost}, \citenamefont
  {Rossberg}, \citenamefont {Dieste-Blanco}, \citenamefont {Raison},
  \citenamefont {Cambriani}, \citenamefont {Somers}, \citenamefont {Bosbach},\
  and\ \citenamefont {Neumeier}}]{Arinicheva:17}%
  \BibitemOpen
  \bibfield  {author} {\bibinfo {author} {\bibfnamefont {Y.}~\bibnamefont
  {Arinicheva}}, \bibinfo {author} {\bibfnamefont {K.}~\bibnamefont {Popa}},
  \bibinfo {author} {\bibfnamefont {A.~C.}\ \bibnamefont {Scheinost}}, \bibinfo
  {author} {\bibfnamefont {A.}~\bibnamefont {Rossberg}}, \bibinfo {author}
  {\bibfnamefont {O.}~\bibnamefont {Dieste-Blanco}}, \bibinfo {author}
  {\bibfnamefont {P.}~\bibnamefont {Raison}}, \bibinfo {author} {\bibfnamefont
  {A.}~\bibnamefont {Cambriani}}, \bibinfo {author} {\bibfnamefont
  {J.}~\bibnamefont {Somers}}, \bibinfo {author} {\bibfnamefont
  {D.}~\bibnamefont {Bosbach}}, \ and\ \bibinfo {author} {\bibfnamefont
  {S.}~\bibnamefont {Neumeier}},\ }\href@noop {} {\bibfield  {journal}
  {\bibinfo  {journal} {Journal of Nuclear Materials}\ }\textbf {\bibinfo
  {volume} {493}},\ \bibinfo {pages} {404 } (\bibinfo {year}
  {2017})}\BibitemShut {NoStop}%
\bibitem [{\citenamefont {Yingjie}\ and\ \citenamefont
  {Vance}(2008)}]{Zhang:08}%
  \BibitemOpen
  \bibfield  {author} {\bibinfo {author} {\bibfnamefont {Z.}~\bibnamefont
  {Yingjie}}\ and\ \bibinfo {author} {\bibfnamefont {E.~R.}\ \bibnamefont
  {Vance}},\ }\href {\doibase DOI:101016/jjnucmat200711011} {\bibfield
  {journal} {\bibinfo  {journal} {Journal of Nuclear Materials}\ }\textbf
  {\bibinfo {volume} {375}},\ \bibinfo {pages} {311} (\bibinfo {year}
  {2008})}\BibitemShut {NoStop}%
\bibitem [{\citenamefont {Lumpkin}\ and\ \citenamefont
  {Geisler-Wierwille}(2012)}]{Lumpkin:12}%
  \BibitemOpen
  \bibfield  {author} {\bibinfo {author} {\bibfnamefont {G.}~\bibnamefont
  {Lumpkin}}\ and\ \bibinfo {author} {\bibfnamefont {T.}~\bibnamefont
  {Geisler-Wierwille}},\ }in\ \href@noop {} {\emph {\bibinfo {booktitle}
  {Comprehensive Nuclear Materials}}},\ \bibinfo {editor} {edited by\ \bibinfo
  {editor} {\bibfnamefont {R.~J.}\ \bibnamefont {Konings}}}\ (\bibinfo
  {publisher} {Elsevier},\ \bibinfo {address} {Oxford},\ \bibinfo {year}
  {2012})\ pp.\ \bibinfo {pages} {563 -- 600}\BibitemShut {NoStop}%
\bibitem [{\citenamefont {Abarenkov}\ and\ \citenamefont
  {Boyko}(2016)}]{Abarenkov:16}%
  \BibitemOpen
  \bibfield  {author} {\bibinfo {author} {\bibfnamefont {I.~V.}\ \bibnamefont
  {Abarenkov}}\ and\ \bibinfo {author} {\bibfnamefont {M.~A.}\ \bibnamefont
  {Boyko}},\ }\href {\doibase DOI: 10.1002/qua.25041} {\bibfield  {journal}
  {\bibinfo  {journal} {Int.\ J.\ Quantum Chem.}\ }\textbf {\bibinfo {volume}
  {116}},\ \bibinfo {pages} {211–236} (\bibinfo {year} {2016})}\BibitemShut
  {NoStop}%
\bibitem [{\citenamefont {Bugaenko}\ \emph {et~al.}(2008)\citenamefont
  {Bugaenko}, \citenamefont {Ryabyh},\ and\ \citenamefont
  {Bugaenko}}]{Bugaenko:08}%
  \BibitemOpen
  \bibfield  {author} {\bibinfo {author} {\bibfnamefont {L.~T.}\ \bibnamefont
  {Bugaenko}}, \bibinfo {author} {\bibfnamefont {S.~M.}\ \bibnamefont
  {Ryabyh}}, \ and\ \bibinfo {author} {\bibfnamefont {A.~L.}\ \bibnamefont
  {Bugaenko}},\ }\href@noop {} {\bibfield  {journal} {\bibinfo  {journal}
  {Vestnik Moscovskogo universiteta, seriya "khimiya"}\ }\textbf {\bibinfo
  {volume} {49}} (\bibinfo {year} {2008})}\BibitemShut {NoStop}%
\bibitem [{\citenamefont {Shakhova}\ \emph {et~al.}(2019)\citenamefont
  {Shakhova}, \citenamefont {Lomachuk}, \citenamefont {Maltsev}, \citenamefont
  {Mosyagin}, \citenamefont {Titov} \emph {et~al.}}]{Shakhova:19a}%
  \BibitemOpen
  \bibfield  {author} {\bibinfo {author} {\bibfnamefont {V.~M.}\ \bibnamefont
  {Shakhova}}, \bibinfo {author} {\bibfnamefont {Y.~V.}\ \bibnamefont
  {Lomachuk}}, \bibinfo {author} {\bibfnamefont {D.~A.}\ \bibnamefont
  {Maltsev}}, \bibinfo {author} {\bibfnamefont {N.~S.}\ \bibnamefont
  {Mosyagin}}, \bibinfo {author} {\bibfnamefont {A.~V.}\ \bibnamefont {Titov}},
   \emph {et~al.},\ }\href@noop {} {\bibfield  {journal} {\bibinfo  {journal}
  {arXiv}\ } (\bibinfo {year} {2019})},\ \bibinfo {note} {to be
  published}\BibitemShut {NoStop}%
\bibitem [{\citenamefont {{van~W\"ullen}}(2010)}]{Wullen:10}%
  \BibitemOpen
  \bibfield  {author} {\bibinfo {author} {\bibfnamefont {C.}~\bibnamefont
  {{van~W\"ullen}}},\ }\href@noop {} {\bibfield  {journal} {\bibinfo  {journal}
  {{Z}.\ {P}hys.\ {C}hem.}\ }\textbf {\bibinfo {volume} {224}},\ \bibinfo
  {pages} {413} (\bibinfo {year} {2010})}\BibitemShut {NoStop}%
\bibitem [{\citenamefont {Adamo}\ and\ \citenamefont
  {Barone}(1999)}]{Adamo:99}%
  \BibitemOpen
  \bibfield  {author} {\bibinfo {author} {\bibfnamefont {C.}~\bibnamefont
  {Adamo}}\ and\ \bibinfo {author} {\bibfnamefont {V.}~\bibnamefont {Barone}},\
  }\href@noop {} {\bibfield  {journal} {\bibinfo  {journal} {J.\ Chem.\ Phys.}\
  }\textbf {\bibinfo {volume} {110}},\ \bibinfo {pages} {6158} (\bibinfo {year}
  {1999})}\BibitemShut {NoStop}%
\bibitem [{\citenamefont {Dovesi}\ \emph {et~al.}(2018)\citenamefont {Dovesi},
  \citenamefont {Erba}, \citenamefont {Orlando}, \citenamefont
  {Zicovich-Wilson}, \citenamefont {Civalleri}, \citenamefont {Maschio},
  \citenamefont {Rérat}, \citenamefont {Casassa}, \citenamefont {Baima},
  \citenamefont {Salustro},\ and\ \citenamefont {Kirtman}}]{Dovesi:18}%
  \BibitemOpen
  \bibfield  {author} {\bibinfo {author} {\bibfnamefont {R.}~\bibnamefont
  {Dovesi}}, \bibinfo {author} {\bibfnamefont {A.}~\bibnamefont {Erba}},
  \bibinfo {author} {\bibfnamefont {R.}~\bibnamefont {Orlando}}, \bibinfo
  {author} {\bibfnamefont {C.~M.}\ \bibnamefont {Zicovich-Wilson}}, \bibinfo
  {author} {\bibfnamefont {B.}~\bibnamefont {Civalleri}}, \bibinfo {author}
  {\bibfnamefont {L.}~\bibnamefont {Maschio}}, \bibinfo {author} {\bibfnamefont
  {M.}~\bibnamefont {Rérat}}, \bibinfo {author} {\bibfnamefont
  {S.}~\bibnamefont {Casassa}}, \bibinfo {author} {\bibfnamefont
  {J.}~\bibnamefont {Baima}}, \bibinfo {author} {\bibfnamefont
  {S.}~\bibnamefont {Salustro}}, \ and\ \bibinfo {author} {\bibfnamefont
  {B.}~\bibnamefont {Kirtman}},\ }\href {\doibase 10.1002/wcms.1360} {\bibfield
   {journal} {\bibinfo  {journal} {Wiley Interdisciplinary Reviews:
  Computational Molecular Science}\ }\textbf {\bibinfo {volume} {8}},\ \bibinfo
  {pages} {1360} (\bibinfo {year} {2018})}\BibitemShut {NoStop}%
\bibitem [{\citenamefont {{URL:
  http://www.qchem.pnpi.spb.ru/Basis/~}}()}]{QCPNPI:Basis}%
  \BibitemOpen
  \bibfield  {author} {\bibinfo {author} {\bibnamefont {{URL:
  http://www.qchem.pnpi.spb.ru/Basis/~}}},\ }\href@noop {} {}\bibinfo {note}
  {~{GRECPs} and basis sets}\BibitemShut {NoStop}%
\bibitem [{\citenamefont {Peintinger}\ \emph {et~al.}(2013)\citenamefont
  {Peintinger}, \citenamefont {Oliveira},\ and\ \citenamefont
  {Bredow}}]{Peintinger:13}%
  \BibitemOpen
  \bibfield  {author} {\bibinfo {author} {\bibfnamefont {M.~F.}\ \bibnamefont
  {Peintinger}}, \bibinfo {author} {\bibfnamefont {D.~V.}\ \bibnamefont
  {Oliveira}}, \ and\ \bibinfo {author} {\bibfnamefont {T.}~\bibnamefont
  {Bredow}},\ }\href@noop {} {\bibfield  {journal} {\bibinfo  {journal}
  {Journal of Computational Chemistry}\ }\textbf {\bibinfo {volume} {34}},\
  \bibinfo {pages} {451} (\bibinfo {year} {2013})}\BibitemShut {NoStop}%
\bibitem [{\citenamefont {Ni}\ \emph {et~al.}(1995)\citenamefont {Ni},
  \citenamefont {Hughes},\ and\ \citenamefont {Mariano}}]{Ni:95}%
  \BibitemOpen
  \bibfield  {author} {\bibinfo {author} {\bibfnamefont {Y.}~\bibnamefont
  {Ni}}, \bibinfo {author} {\bibfnamefont {J.~M.}\ \bibnamefont {Hughes}}, \
  and\ \bibinfo {author} {\bibfnamefont {A.~N.}\ \bibnamefont {Mariano}},\
  }\href@noop {} {\bibfield  {journal} {\bibinfo  {journal} {American
  Mineralogist}\ }\textbf {\bibinfo {volume} {80}},\ \bibinfo {pages} {21}
  (\bibinfo {year} {1995})}\BibitemShut {NoStop}%
\bibitem [{\citenamefont {Tang}\ \emph {et~al.}(2009)\citenamefont {Tang},
  \citenamefont {Sanville},\ and\ \citenamefont {Henkelman}}]{Tang:09}%
  \BibitemOpen
  \bibfield  {author} {\bibinfo {author} {\bibfnamefont {W.}~\bibnamefont
  {Tang}}, \bibinfo {author} {\bibfnamefont {E.}~\bibnamefont {Sanville}}, \
  and\ \bibinfo {author} {\bibfnamefont {G.}~\bibnamefont {Henkelman}},\
  }\href@noop {} {\bibfield  {journal} {\bibinfo  {journal} {J.\ Phys.:
  Condensed Matter}\ }\textbf {\bibinfo {volume} {21}},\ \bibinfo {pages}
  {084204} (\bibinfo {year} {2009})}\BibitemShut {NoStop}%
\bibitem [{\citenamefont {Bader}(1998)}]{Bader:98}%
  \BibitemOpen
  \bibfield  {author} {\bibinfo {author} {\bibfnamefont {R.~F.~W.}\
  \bibnamefont {Bader}},\ }in\ \href@noop {} {\emph {\bibinfo {booktitle}
  {Encyclopedia of Computational Chemistry}}},\ Vol.~\bibinfo {volume} {1}\
  (\bibinfo  {publisher} {Wiley},\ \bibinfo {address} {Chichester, U.K.},\
  \bibinfo {year} {1998})\ pp.\ \bibinfo {pages} {64--86}\BibitemShut {NoStop}%
\bibitem [{\citenamefont {Titov}\ \emph {et~al.}(2014)\citenamefont {Titov},
  \citenamefont {Lomachuk},\ and\ \citenamefont {Skripnikov}}]{Titov:14a}%
  \BibitemOpen
  \bibfield  {author} {\bibinfo {author} {\bibfnamefont {A.~V.}\ \bibnamefont
  {Titov}}, \bibinfo {author} {\bibfnamefont {Y.~V.}\ \bibnamefont {Lomachuk}},
  \ and\ \bibinfo {author} {\bibfnamefont {L.~V.}\ \bibnamefont {Skripnikov}},\
  }\href@noop {} {\bibfield  {journal} {\bibinfo  {journal} {Phys.\ Rev.\ A}\
  }\textbf {\bibinfo {volume} {90}},\ \bibinfo {pages} {052522} (\bibinfo
  {year} {2014})}\BibitemShut {NoStop}%
\bibitem [{\citenamefont {Lomachuk}\ and\ \citenamefont
  {Titov}(2013)}]{Lomachuk:13}%
  \BibitemOpen
  \bibfield  {author} {\bibinfo {author} {\bibfnamefont {Y.~V.}\ \bibnamefont
  {Lomachuk}}\ and\ \bibinfo {author} {\bibfnamefont {A.~V.}\ \bibnamefont
  {Titov}},\ }\href@noop {} {\bibfield  {journal} {\bibinfo  {journal} {Phys.\
  Rev.\ A}\ }\textbf {\bibinfo {volume} {88}},\ \bibinfo {pages} {062511}
  (\bibinfo {year} {2013})}\BibitemShut {NoStop}%
\bibitem [{\citenamefont {Lomachuk}\ \emph {et~al.}(2018)\citenamefont
  {Lomachuk}, \citenamefont {Demidov}, \citenamefont {Skripnikov},
  \citenamefont {Zaitsevskii}, \citenamefont {Semenov}, \citenamefont
  {Mosyagin},\ and\ \citenamefont {Titov}}]{Lomachuk:18en}%
  \BibitemOpen
  \bibfield  {author} {\bibinfo {author} {\bibfnamefont {Y.~V.}\ \bibnamefont
  {Lomachuk}}, \bibinfo {author} {\bibfnamefont {Y.~A.}\ \bibnamefont
  {Demidov}}, \bibinfo {author} {\bibfnamefont {L.~V.}\ \bibnamefont
  {Skripnikov}}, \bibinfo {author} {\bibfnamefont {A.~V.}\ \bibnamefont
  {Zaitsevskii}}, \bibinfo {author} {\bibfnamefont {S.~G.}\ \bibnamefont
  {Semenov}}, \bibinfo {author} {\bibfnamefont {N.~S.}\ \bibnamefont
  {Mosyagin}}, \ and\ \bibinfo {author} {\bibfnamefont {A.~V.}\ \bibnamefont
  {Titov}},\ }\href@noop {} {\bibfield  {journal} {\bibinfo  {journal} {Optics
  and Spectroscopy}\ }\textbf {\bibinfo {volume} {124}},\ \bibinfo {pages}
  {472} (\bibinfo {year} {2018})}\BibitemShut {NoStop}%
\bibitem [{\citenamefont {Tupitsyn}\ \emph {et~al.}(2002)\citenamefont
  {Tupitsyn}, \citenamefont {Deyneka},\ and\ \citenamefont {Bratzev}}]{hfd}%
  \BibitemOpen
  \bibfield  {author} {\bibinfo {author} {\bibfnamefont {I.~I.}\ \bibnamefont
  {Tupitsyn}}, \bibinfo {author} {\bibfnamefont {G.~B.}\ \bibnamefont
  {Deyneka}}, \ and\ \bibinfo {author} {\bibfnamefont {V.~F.}\ \bibnamefont
  {Bratzev}},\ }\href@noop {} {\enquote {\bibinfo {title} {{``{\sc hfd}''}},}\
  } (\bibinfo {year} {1977--2002}),\ \bibinfo {note} {{\sc hfd}, a program for
  atomic finite-difference four-component {D}irac-{H}artree-{F}ock calculations
  on the base of the HFD code~\cite{Bratzev:77}}\BibitemShut {NoStop}%
\bibitem [{\citenamefont {Bratzev}\ \emph {et~al.}(1977)\citenamefont
  {Bratzev}, \citenamefont {Deyneka},\ and\ \citenamefont
  {Tupitsyn}}]{Bratzev:77}%
  \BibitemOpen
  \bibfield  {author} {\bibinfo {author} {\bibfnamefont {V.~F.}\ \bibnamefont
  {Bratzev}}, \bibinfo {author} {\bibfnamefont {G.~B.}\ \bibnamefont
  {Deyneka}}, \ and\ \bibinfo {author} {\bibfnamefont {I.~I.}\ \bibnamefont
  {Tupitsyn}},\ }\href@noop {} {\bibfield  {journal} {\bibinfo  {journal}
  {Bull.\ Acad.\ Sci.\ USSR, Phys.\ Ser.}\ }\textbf {\bibinfo {volume} {41}},\
  \bibinfo {pages} {173} (\bibinfo {year} {1977})}\BibitemShut {NoStop}%
\end{thebibliography}
\end{document}